УДК 378.147　　　　　　　　　　　　МАРКОВА О. М.,
старший викладач кафедри комп'ютерних систем та мереж ДВНЗ «Криворізький національний університет»

## МОДЕЛЬ МЕТОДИЧНОЇ СИСТЕМИ ТА ЦІЛІ НАВЧАННЯ ОСНОВ МАТЕМАТИЧНОЇ ІНФОРМАТИКИ СТУДЕНТІВ ТЕХНІЧНИХ УНІВЕРСИТЕТІВ

*У статті наведені етапи розробки моделі методичної системи навчання спецкурсу «Основи математичної інформатики» студентів технічних університетів. Конкретизовано цільовий компонент методичної системи.*

***Ключові слова:*** *модель, методична система, принципи навчання, цілі навчання.*

**Постановка проблеми.** Розробка методичної системи навчання спецкурсу «Основи математичної інформатики» відіграє ключову роль у формуванні компетентностей з математичної інформатики студентів технічних університетів, оскільки лише у ньому відбувається комплексне формування всіх їх компонентів [9]. Тому актуальним є аналіз складників методичної системи, виявлення найбільш слабких місць і проблем, що здатні помітно погіршити її якості і без подолання яких неможливий її подальший розвиток.

**Аналіз останніх досліджень і публікацій.** Традиційною моделлю методичної системи навчання є п'ятикомпонентна модель, запропонована А. М. Пишкало [9], в якій використовується системний підхід стосовно компонентів процесу навчання (всі компоненти утворюють єдине ціле із визначеними внутрішніми зв'язками). Згідно з цією моделлю, методична система навчання – це сукупність ієрархічно пов'язаних компонентів: цілей навчання, змісту, методів, засобів і форм організації навчання (рис. 1). Функціонування методичної системи підпорядковано закономірностям, що пов'язані з внутрішньою будовою самої системи, коли зміна однієї чи декількох її компонентів призведе до зміни всієї системи.

У зміст та обсяг поняття «методична система» явно чи неявно включаються принципи навчання. У моделі А. М. Пишкало, на відміну від моделі Ч. Купісевича, вони є неявними: дослідник вказує, що співвіднесення принципів із цілями, змістом, засобами або процесом навчання призводить до втрати ними ознак загальних норм дидактичної діяльності [2, с. 147-152]. У зв'язку з цим принципи навчання при побудові методичних систем ураховуються (табл. 1), але до відповідних моделей не включаються.

**Таблиця 1**
*Співвідношення основних компонентів навчального процесу та принципів навчання (за В. А. Байдаком [1, с. 31])*

| Основні компоненти навчального процесу | Принципи навчання |
|---|---|
| Задачі навчання | Принцип спрямованості навчання на рішення у взаємозв'язку задач освіти, виховання та загального розвитку тих, хто навчається |
| Зміст навчання | Принципи: науковості навчання; зв'язку навчання з життям, практикою; систематичності та послідовності навчання; доступності |
| Методи навчання та відповідні їм засоби | Принципи: наочності навчання; свідомості та активності тих, хто навчається, за керівної ролі викладача; поєднання різних методів, а також засобів навчання у залежності від задач та змісту навчання |
| Форми організації навчання | Принцип поєднання різноманітних форм навчання в залежності від задач, змісту та методів навчання |
| Умови навчання | Принцип створення необхідних умов для навчання |
| Результати навчання | Принцип міцності, усвідомленості та дієвості результатів навчання, виховання та розвитку |

Розглядаючи сукупність таких компонентів традиційної методичної системи навчання, як методи, форми організації та засоби навчання, услід за Л. О. Черних вважаємо, що вони утворюють певну підсистему єдиної системи, що називають *технологією навчання* [10]. Схематичне зображення структури методичної системи навчання з виділеною пунктиром технологічною підсистемою подано на рис. 2.

Виокремлення технології навчання з методичної системи навчання зумовлено суттєво більш тісними зв'язками між її компонентами: адже «підсумком теоретичного узагальнення педагогічного та методичного матеріалу» [9, с. 42] була структура методичної системи, у якій цілі та зміст навчання впливали на технологічні складові, як це показано у [1, с. 25].

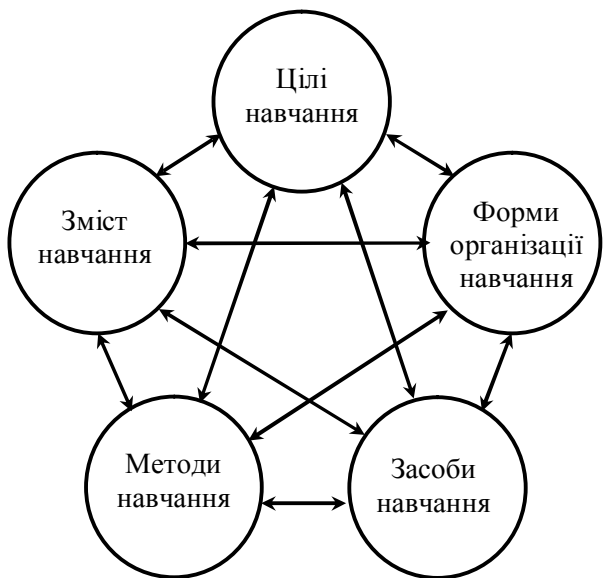 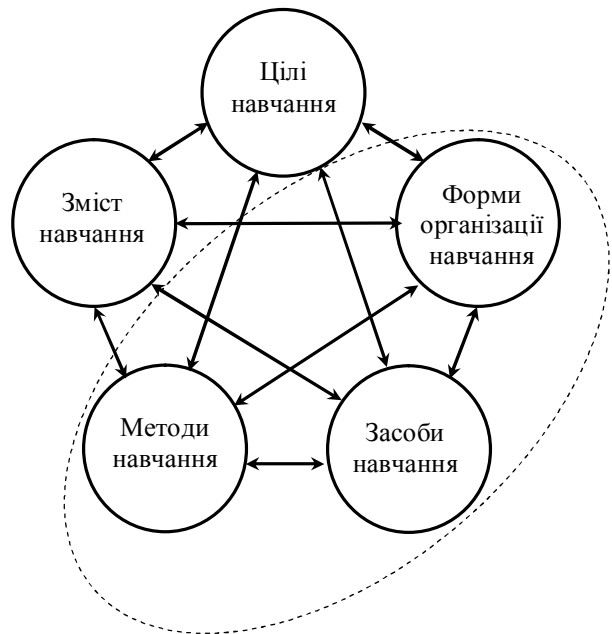

*Рис. 1. Класична структура методичної системи навчання (за А. М. Пишкало)*

*Рис. 2. Структура методичної системи навчання з виділеною підсистемою «технологія навчання» (за Л. О. Черних)*

Інший підхід пропонують В. В. Лаптєв, Н. І. Рижова, М. В. Швецький, які об'єднують методи, форми організації, засоби та зміст навчання у методичне забезпечення, виводячи цілі навчання за межі методичної системи та доповнюючи її очікуваними результатами навчання і трьома технологіями: відбору методів, форм та засобів навчання; відбору змісту навчання; встановлення зв'язків між елементами системи [3, с. 109-110].

«Можна говорити про те, що поява принципово нових засобів навчання, що якісно змінюють можливості передавання інформації і розширюють можливості організації навчального процесу, приводить до перегляду змісту, форм і методів навчання і може опосередковано позначитися на цілях навчання» [7, с. 7]. Це зауваження майже на 10 років випередило появу комп'ютерів у масовій школі, але з позицій сьогодення можна стверджувати, що в ньому сконцентровані всі основні ідеї створення й обґрунтування методичної системи навчання спецкурсу «Основи математичної інформатики»: використання ІКТ у цілому та хмарних технологій зокрема як засобу навчання в значній мірі обумовлює цілі, зміст, методи й форми організації навчання в сучасному технічному університеті.

**Мета статті.** Розробити модель методичної системи навчання спецкурсу «Основи математичної інформатики» студентів технічних університетів та конкретизувати її цільовий компонент.

**Виклад основного матеріалу.** Виділення засобів хмарних технологій навчання основ математичної інформатики вимагає побудови технології навчання, зумовлюючи вибір відповідних хмаро орієнтованих форм організації та методів навчання. З іншого боку, теорія, методи та засоби хмарних технологій [8] суттєво впливають на основний зміст навчання [5] та його цілі. Таким чином, теорія, методи та засоби хмарних технологій є основою побудови моделі методичної системи навчання спецкурсу «Основи математичної інформатики» (рис. 3).

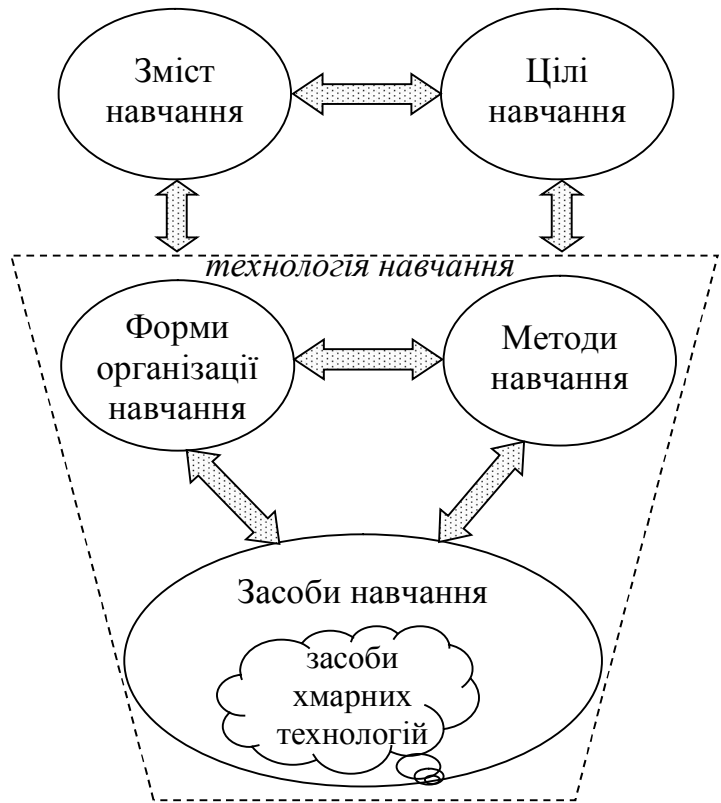

*Рис. 3. Модель методичної системи навчання основ математичної інформатики*

Виділення в структурі методичної системи навчання спецкурсу «Основи математичної інформатики» технологічної підсистеми надало можливість максимально відобразити взаємовпливи всіх її компонентів: цільового, змістового та технологічного.

При створенні методичної системи навчання спецкурсу «Основи математичної інформатики» було необхідно:

– урахувати професійну спрямованість підготовки майбутніх фахівців з інформаційних технологій шляхом відбору змісту навчання та професійно орієнтованих засобів хмарних технологій;

– забезпечити фундаментальність навчання шляхом системного опанування та застосування моделей і методів математичної інформатики;

– спрогнозувати результати педагогічного впливу, передбачаючи, які професійні компетентності мають бути сформовані у процесі навчання основ математичної інформатики та як будуть використовуватись засоби хмарних технологій у подальшій професійній підготовці.

Отже, виходячи з визначеної структури, необхідно виділити насамперед цільовий компонент методичної системи навчання спецкурсу «Основи математичної інформатики».

Мета (ціль) навчання – ідеалізоване передбачення кінцевих результатів навчання; те, до чого прагнуть учасники навчального процесу – студенти і викладачі. Виділяють три основні групи взаємопов'язаних цілей: 1) освітня – формування у студентів наукових знань, загальнонавчальних і спеціальних умінь та навичок; 2) розвивальна – розвиток мови, мислення, пам'яті, здібностей, рухової та сенсорної систем тощо; 3) виховна – формування світогляду, моралі, естетичної культури тощо.

Загальні цілі навчання інформатики визначаються специфікою її внеску в розв'язання основних задач загальної освіти людини: формування основ наукового світогляду, що полягає у розумінні сучасної картини світу, зміні змісту та характеру діяльності людини відповідно до розвитку нових інформаційних технологій; розвиток мислення тих, хто навчається, зокрема теоретичного, творчого та мислення нового типу (операційного) у процесі вивчення інформатики з використанням комп'ютерної техніки; підготовка тих, хто навчається, до практичної діяльності та продовження освіти, оскільки саме вивчення інформатики впливає на формування комп'ютерної грамотності та інформаційної культури, що уможливлює існування людини у сучасному інформаційному суспільстві [3, с. 268].

У процесі здобуття загальної середньої освіти елементи математичної інформатики реалізуються насамперед в інформаційно-комунікаційному компоненті відповідного стандарту через формування в учнів навичок і вмінь проводити основні операції з інформаційними об'єктами, зокрема:

– створювати інформаційні об'єкти, фіксувати, записувати, спостерігати за ними і вимірювати їх, зокрема, в рамках реалізації індивідуальних і колективних проектів;

– висувати і перевіряти нескладні гіпотези навчально-пізнавального характеру, створювати, вивчати та використовувати інформаційні об'єкти;

– вивчати, аналізувати інформаційні процеси, що відбуваються у живій природі, суспільстві та техніці;

– одержувати уявлення про основи управління, прийняття рішень, основні принципи роботи засобів інформаційних технологій;

– виявляти та аналізувати інформаційні процеси в технічних, біологічних і соціальних системах;

– будувати і використовувати інформаційні моделі, а також засоби опису та моделювання явищ і процесів [8].

Загальною метою навчання інформатики студентів технічних університетів є:

1) формування уявлень про ідеї та методи дискретної математики як форми опису та подання об'єктів навколишнього світу;

2) формування та розвиток операційного стилю мислення майбутніх ІТ-фахівців (вміння формалізувати задачу; виділити в ній логічно самостійні частини; визначити взаємозв'язок цих частин; спроектувати рішення за допомогою спадної чи висхідної технологій; верифікувати результат);

3) формування уявлення про методологію інформатики (обчислювальний експеримент) як основний метод пізнання в галузі інформатики.

Виходячи з даного трактування, *мету навчання* спецкурсу «Основи математичної інформатики» студентів технічних університетів визначимо через формування компетентностей з математичної інформатики.

*Цілі навчання* спецкурсу «Основи математичної інформатики» окреслені такими завданнями:

– ознайомлення зі структурами даних і алгоритмами, які є фундаментом сучасної методології розробки програм;

– вивчення методів розв'язування інженерних та наукових задач з використанням чисельних методів;

– ознайомлення з основними принципами кодування і модуляції сигналів у процесі передавання даних, опрацювання сигналів, збільшення перешкодозахищеності при передаванні даних по каналах зв'язку;

– формування умінь описувати основні методи реєстрації сигналів, декодування і виявлення помилок за допомогою різних коригуючих кодів;

– вивчення основ алгоритмічних аспектів теорії чисел та їх застосування в сучасній криптографії;

– опанування хмарними технологіями для практичної реалізації основних методів математичної інформатики.

**Висновки.** 1. Розроблена модель методичної системи навчання спецкурсу «Основи математичної інформатики» містить у якості компонентів зміст, цілі та технологію навчання. До складу останньої входять форми організації, методи та засоби навчання, серед яких провідними є засоби хмарних технологій.

2. Основною метою навчання спецкурсу «Основи математичної інформатики» є формування компетентностей студентів технічних університетів з математичної інформатики.

**MARKOVA O.**,
Senior Lecturer of Computer Systems and Networks Department, SIHE «Kryvyi Rih National University»

**THE MODEL OF METHODICAL SYSTEM AND LEARNING OBJECTIVES OF THE FOUNDATIONS OF MATHEMATICAL INFORMATICS FOR STUDENTS OF TECHNICAL UNIVERSITIES**

***Abstract. Introduction.*** *Development of methodical system for training course "Foundations of mathematical informatics" plays a key role in forming the students' of technical universities competencies in mathematical informatics. So it is very important to analyze of the components of methodical system, identify the weaknesses and problems that can significantly impair its quality and which can not be overcome without its further development.*

***Purpose.*** *Develop a model of methodical system of training course "Foundations of mathematical informatics" for students of technical universities and specify its target component.*

***Methods.*** *Using cloud technologies at learning the foundations of mathematical informatics requires the construction technology training, which results in the selection of appropriate cloud-oriented forms of organization and teaching methods. On the other hand, the theory, methods and tools for cloud significantly affect the primary content of learning and its goals. Thus, the cloud technology theory, methods and tools are the basis for constructing methodical system of training course "Foundations of mathematical science".*

***Results.*** *Development of methodical system of training course "Foundations of mathematical informatics" plays a leading role in forming the students' of technical universities competencies in mathematical informatics due to its fundamental impact and technology improvement.*

***Originality.*** *Theoretically grounded and constructed the model of methodical system of training course "Foundations of mathematical informatics" and learning goals.*

***Conclusion.*** *The model of methodical system of training course "Foundations of mathematical informatics" includes content, objectives and learning technology. The last one contains forms of organization, methods and teaching tools, including leading cloud technologies. The main purpose of training course "Foundations of mathematical informatics" is developing of students competencies of technical universities in mathematical informatics.*

***Keywords****: model, methodical system, principles of teaching, learning objectives.*